\documentclass{appolb}
\usepackage{epsfig}
\title{TRUNCATED MOMENTS OF NONSINGLET PARTON DISTRIBUTIONS IN THE DOUBLE
LOGARITHMIC $ln^2x$ APPROXIMATION.}

\author{Dorota Kotlorz $^1$, Andrzej Kotlorz $^2$
\address{$^1$Department of Physics Ozimska 75, $^2$Department of 
Mathematics Luboszycka 3, Technical University of Opole, 
45-370 Opole, Poland, e-mail $^1$: {\tt dstrozik@po.opole.pl}}}

\begin{document}
\pagestyle{plain}
\eqsec
\maketitle

\begin{abstract}
The method of truncated Mellin moments in a solving QCD evolution equations 
of the nonsinglet structure functions $F_2^{NS}(x,Q^2)$ and $g_1^{NS}(x,Q^2)$ 
is presented. All calculations are performed within double logarithmic $ln^2x$ 
approximation. An equation for truncated moments which incorporates $ln^2x$ 
effects is formulated and solved for the unintegrated structure function 
$f^{NS}(x,Q^2)$. The contribution to the Bjorken sum rule coming from the
region of very small $x$ is quantified. Further possible improvement of this 
approach is also discussed.
\end{abstract}


\section{Introduction}

Structure functions play a central role in the perturbative QCD.
Experimental measurements of the spin dependent and unpolarized structure
functions of the nucleon allow verification of sum rules and determination
of free parameters of the input parton distributions. From the other side,
theoretical analysis within perturbative methods investigates the available
experimentally region of the variables $x$ and $Q^2$ and the interesting
very small $x$ region (still unmeasurable) as well. In this low $x$ region
QCD predicts a strong growth of structure functions with decreasing $x$ -
the longitudinal momentum fraction of a hadron carried by a parton. Small
$x$ behaviour of the singlet unpolarized structure functions, driven by
gluons, is governed by BFKL \cite{b1} or CCFM \cite{b2}, \cite{b3} equations which
generate the steep $x^{-\lambda}$ ($\lambda\sim 0.3$) shape of
$F_2^S(x,Q^2)$ \cite{b4}. For the nonsinglet unpolarized $F_2^{NS}(x,Q^2)$
the driving term at small $x$ is a nonperturbative contribution of the
$A_2$ Regge pole $F_2^{NS}\sim x^{0.5}$ \cite{b4} which dominates even over the
$\alpha_s ln^2x$ effects. These double logarithmic terms control however
the small $x$ behaviour of both nonsinglet and singlet spin dependent
structure function $g_1(x,Q^2)$ \cite{b4}, \cite{b5}. Knowledge about structure
functions at very low $x$ is very important. The sum rules, which can be
verified by experiments, concern moments of structure functions
$\int\limits_{0}^{1} dx x^{n-1} g_1(x,Q^2)$ and $\int\limits_{0}^{1} dx x^{n-1}
F_2(x,Q^2)$ and hence require knowledge of $g_1$ and $F_2$ over the entire
region of $x\in (0;1)$. The lowest limit of $x$ in present experiments is
about $x\sim 10^{-5}$ so in theoretical analysis one should extrapolate
results to $x=0$ and $x=1$. More important is however the extrapolation to 
$x\rightarrow 0$, where structure functions grow strongly than the
extrapolation to $x=1$, where structure functions are equal to 0. The limit 
$x\rightarrow 0$ which implies that the invariant energy $W^2$ of the
inelastic lepton-hadron scattering becomes infinite
($W^2=Q^2(\frac{1}{x}-1)$) will never be attained experimentally. So we will
really never know "what it happens" with structure functions at 
$x\rightarrow 0$. This situation is however not quite hopeless. One can
combine the QCD perturbative analysis in the very interesting small $x$
region with experimental data without uncertainty from the region where 
$x\rightarrow 0$. It could be achieved through dealing with truncated
moments of the structure functions, where one takes an integral over
$x_0\leq x \leq 1$ instead over the whole region $0\leq x \leq 1$. In usually
used method of solving evolution equations in QCD one takes Mellin (full)
transforms of these equations what gives possibility of analytical
solutions. Then after inverse Mellin transform (performed numerically) one
can obtain suitable solutions of original equations in $x$ space. In this
way e.g. in a case of DGLAP approximation, the differentio-integral
equations for parton distributions $q(x,Q^2)$ after Mellin transform change
into simple differential and diagonalized ones in moment space $n$. The only
problem is knowledge of input parametrizations for the whole region 
$0\leq x \leq 1$ what is necessary in the determination of moments of
distribution functions. Using truncated moments approach one can avoid
uncertainty from the unmeasurable $x\rightarrow 0$ region and also obtain
important theoretical results incorporating perturbative QCD effects at
small $x$, which could be verified experimentally. Truncated moments of
parton distributions in solving DGLAP equations have been presented in
\cite{b6}. Authors have shown that the evolution equations for truncated
moments though not diagonal can be solved with good precision. This is
because each $n$-th truncated moment couples only with $n+j$ -th ($j\geq 0$)
truncated moments. In our paper we adopt the truncated moments method to
double logarithmic $ln^2x$ resummation. However a technique we use in this
approach is quite different because suitable integral equations which 
correspond to $ln^2x$ terms are of course different from the DGLAP ones. As 
a result we obtain the equations for truncated moments of the unintegrated 
structure function $f^{NS}(x,Q^2)$, where each $n$-th moment ($n\neq 0$) 
couples only with itself and with $0$-th moment. For fixed coupling constant 
$\alpha_s$ the result for $n$-th truncated moment can be found analytically. 
The purpose of this paper is to start the truncated moments method in the case
of perturbative QCD formalism, describing double logarithmic $ln^2x$ terms
in the nonsinglet structure functions $F_2^{NS}$ and $g_1^{NS}$. In the next
section we recall the approach which resums the double logarithmic terms.
The integral equation for the nonsinglet unintegrated quark distributions
$f^{NS}(x,Q^2)$ is presented. In point 3 the equation for truncated moments
of $f^{NS}(x,Q^2)$ within $ln^2x$ approach is derived. This equation is
solved analytically for fixed $\alpha_s$. Agreement for the limit case
$x_0\rightarrow 0$ (full moments) of our results is shown. Results for
truncated moments of $F_2^{NS}$ and $g_1^{NS}$ for simple Regge-type input
parametrizations and different $x_0$ are presented in section 4. We
calculate also the contribution to the Bjorken sum rule coming from the
region of very small $x$. Finally in section 5 we summarize our results and 
discuss further possible improvement of our treatment.

\section{Idea of double logarithmic $ln^2x$ resummation for the nonsinglet
unpolarized and polarized structure functions of the nucleon.}

It has been noticed \cite{b4}, \cite{b5} that the nonsinglet structure
functions of the nucleon in the small $x$ region are governed by double
logarithmic terms i.e. powers of $\alpha_s ln^2x$ at each order of the
perturbative expansion. This contribution to the $ln^2x$ resummation comes
from the ladder diagram with quark and gluon exchanges along the chain -
cf. Fig.1.
\begin{figure}[ht]
\begin{center}
\includegraphics[width=70mm]{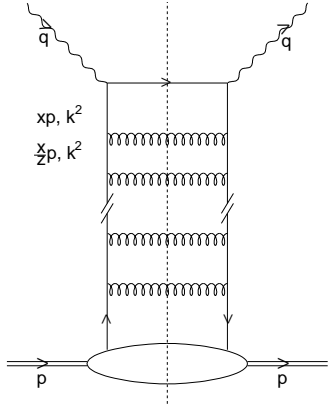}
\caption{A ladder diagram generating double logarithmic $\ln^2(1/x)$
terms in the nonsinglet spin structure functions.}
\end{center}
\end{figure}
In contrast to the singlet spin dependent structure function, for the
nonsinglet one the "bremsstrahlung" nonladder corrections vanish for the
unpolarized structure function and are negligible for the spin dependent one
\cite{b5}, \cite{b7}. In this way we do not need to take into account the
nonladder diagrams in the case of nonsinglet (both polarized and unpolarized)
structure functions. The Regge theory \cite{b8}, which concerns the Regge
limit: $x\rightarrow 0$ predicts singular behaviour of nonsinglet,
unpolarized distributions and nonsingular (flat) shape of nonsinglet
polarized ones:
\begin{equation}\label{r2.1}
q^{NS}\sim x^{-0.5}
\end{equation}
\begin{equation}\label{r2.2}
\Delta q^{NS}\sim x^{0} \div x^{-0.5}
\end{equation}
where $q^{NS}$ denotes nonsinglet unpolarized and $\Delta q^{NS}$ nonsinglet
polarized quark distributions. The nonsinglet part of the unpolarized
structure function has the form:
\begin{equation}\label{r2.3}
F_2^{NS}(x,Q^2) = F_2^{p}(x,Q^2)-F_2^{n}(x,Q^2)
\end{equation}
where
\begin{equation}\label{r2.4}
F_2(x,Q^2) = \sum\limits_{i=u,d,s,...} e_i^2 [xq_i(x,Q^2) +
x\bar{q_i}(x,Q^2)]
\end{equation}
The spin dependent structure function is:
\begin{equation}\label{r2.5}
g_1^{NS}(x,Q^2) = g_1^{p}(x,Q^2)-g_1^{n}(x,Q^2)
\end{equation}
and
\begin{equation}\label{r2.6}
g_1(x,Q^2) = \frac{1}{2}\sum\limits_{i=u,d,s,...} e_i^2 \Delta q_i(x,Q^2)
\end{equation}
$p$ and $n$ in above formulae denote proton and neutron respectively, $e_i$
is a charge of the $i$-flavour quark. Hence finally we get:
\begin{equation}\label{r2.7}
xq^{NS}=F_2^{NS}(x,Q^2) = \frac{x}{3}(u_{val}-d_{val})(x,Q^2)
\end{equation}
and
\begin{equation}\label{r2.8}
\Delta q^{NS}=g_1^{NS}(x,Q^2) = \frac{1}{6}(\Delta u_{val}-
\Delta d_{val})(x,Q^2)
\end{equation}
where $u_{val}$, $d_{val}$, $\Delta u_{val}$, $\Delta d_{val}$ are
respectively spin and nonspin valence quark distributions in the proton.
In the double logarithmic approximation the unintegrated nonsinglet quark
distribution function $f^{NS}(x,k^2)$ satisfies the following integral
equation \cite{b4}:
\begin{equation}\label{r2.9}
f^{NS}(x,k^2)=f_0^{NS}(x)+\bar{\alpha_s}
\int\limits_x^1\frac{dz}{z}\int\limits_{k_0^2}^{k^2/z}
\frac{dk'^2}{k'^2}f^{NS}(\frac{x}{z},k'^2)
\end{equation}
where
\begin{equation}\label{r2.10}
\bar{\alpha_s}=\frac{2\alpha_s}{3\pi}
\end{equation}
and $f_0^{NS}(x)$ is a nonperturbative contribution which has a form:
\begin{equation}\label{r2.11}
f_0^{NS}(x)=\bar{\alpha_s}\int\limits_x^1\frac{dz}{z}
q^{NS}(z)\sim \bar{\alpha_s} x^{-0.5}
\end{equation}
for nonspin distributions or
\begin{equation}\label{r2.12}
f_0^{NS}(x)=\bar{\alpha_s}\int\limits_x^1\frac{dz}{z}
\Delta q^{NS}(z)\sim \bar{\alpha_s} \ln\frac{1}{x}
\end{equation}
for spin dependent distributions. The driving terms of these nonperturbative
contributions $q^{NS}$ and $\Delta q^{NS}$ are shown in (\ref{r2.1}) and
(\ref{r2.2}). The unintegrated distribution $f^{NS}(x,k^2)$ is related to
the quark distributions $q^{NS}$ ($\Delta q^{NS}$) via:
\begin{equation}\label{r2.13}
f^{NS}(x,k^2)=\frac{\partial\frac{1}{x} F_2^{NS}(x,k^2)}{\partial\ln k^2}
\end{equation}
in the unpolarized case and
\begin{equation}\label{r2.14}
f^{NS}(x,k^2)=\frac{\partial g_1^{NS}(x,k^2)}{\partial\ln k^2}
\end{equation}
in the polarized one. Using a method of the Mellin moment functions one can 
obtain from (\ref{r2.9}) the following equation:
\begin{equation}\label{r2.15}
\bar{f}^{NS}(n,k^2)=\bar{f_0}^{NS}(n)+\frac {\bar{\alpha_s}}{n}
\left[\int\limits_{k_0^2}^{k^2}\frac{dk'^2}{k'^2}\bar{f}^{NS}(n,k'^2)+
\int\limits_{k^2}^{\infty}\frac{dk'^2}{k'^2} \left(\frac{k^2}{k'^2}\right)^n
\bar{f}^{NS}(n,k'^2)\right]
\end{equation}
where the full Mellin moment of a function $f(x)$ is defined as:
\begin{equation}\label{r2.16}
{\it M}(n)\equiv\int\limits_0^1 dx x^{n-1} f(x)
\end{equation}
and in our case:
\begin{equation}\label{r2.17}
\bar{f}^{NS}(n,k^2)=\int\limits_0^1 dx x^{n-1} f^{NS}(x,k^2)
\end{equation}
As it was shown in \cite{b4}, equation (\ref{r2.15}) for fixed coupling
$\bar{\alpha_s}$ can be solved analytically. Because we would like later to
compare full-moments approach with the truncated moments one, let us recall
the way to obtain the analytical solution of (\ref{r2.15}). A short
explanation of this is given in Appendix A. Thus we get the solution of
(\ref{r2.15}) in the form:
\begin{equation}\label{r2.18}
\bar{f}^{NS}(n,k^2) = 
\bar{f_0}^{NS}(n)\frac{n\gamma}{\bar{\alpha_s}}
\left(\frac{k^2}{k_0^2}\right)^{\gamma}
\end{equation}
where
\begin{equation}\label{r2.19}
\gamma = \frac{n}{2}\left[1-\sqrt{1 - (\frac{n_0}{n})^2}\right]
\end{equation}
and
\begin{equation}\label{r2.20}
n_0 = 2\sqrt{\bar{\alpha_s}}
\end{equation}
The inhomogeneous term $\bar{f_0}^{NS}(n)$ in (\ref{r2.15}) and
(\ref{r2.18}) according to (\ref{r2.16}), (\ref{r2.11})-(\ref{r2.12}) and
(\ref{r2.1})-(\ref{r2.2}) behaves as:
\begin{equation}\label{r2.21}
\bar{f_0}^{NS}(n) \sim \frac{\bar{\alpha_s}}{n(n-0.5)}
\end{equation}
for unpolarized structure functions and
\begin{equation}\label{r2.22}
\bar{f_0}^{NS}(n) \sim \frac{\bar{\alpha_s}}{n^2}
\end{equation}
for the spin dependent ones. The anomalous dimension of the moment of the
nonsinglet structure function $\gamma$ from (\ref{r2.19}) has a (square
root) branch point singularity at $n=n_0$. This gives the following
behaviour of the nonsinglet structure functions at small $x$:
\begin{equation}\label{r2.23}
f^{NS}(x,k^2) \sim x^{-n_0} \left(\frac{k^2}{k_0^2}\right)^{n_0/2}
\end{equation}
and hence also
\begin{equation}\label{r2.24}
\frac{1}{x} F_2^{NS}(x,k^2) \sim g_1^{NS}(x,k^2) 
\sim x^{-n_0} \left(\frac{k^2}{k_0^2}\right)^{n_0/2}
\end{equation}
This low $x$ shape $\sim x^{-n_0}$, where $n_0$ given in (\ref{r2.20}) is
equal to 0.39 for $\bar{\alpha_s} = 0.038$ ($\alpha_s = 0.18$), remains
nonleading in the case of the nonsinglet unpolarized structure function in
comparison to the contribution of the nonperturbative $A_2$ Regge pole
(\ref{r2.1}). In this way QCD perturbative singularity at small $x$
generated by the double logarithmic $ln^2x$ resummation for nonsinglet
unpolarized quark distributions is hidden behind the leading Regge
contribution:
\begin{equation}\label{r2.25}
q^{NS}(x,Q^2)\sim x^{-n_0} < x^{-\alpha_{A_2}(0)}~~~\alpha_{A_2}(0) = 0.5
\end{equation}
Quite different situation occurs for the nonsinglet spin dependent
functions, where the double logarithmic contribution becomes important:
\begin{equation}\label{r2.26}
\Delta q^{NS}(x,Q^2)\sim x^{-n_0} > x^{-\alpha_{A_1}(0)}~~~\alpha_{A_1}(0)
\leq 0
\end{equation}
This takes place because the nonperturbative Regge part for spin dependent
quark distributions involves a very low intercept $\alpha_{A_1}(0)\leq 0$.
Small $x$ behaviour $\sim x^{-2\sqrt{\bar{\alpha_s}}}$ of the nonsinglet
structure functions originating from double logarithmic $ln^2x$ resummation 
is a very interesting feature. Particularly for the polarized structure
functions, where the double logarithmic analysis enable estimation of parton
parametrizations at low $x$. In the next section we introduce the truncated
moments method and combine it with the $ln^2x$ approach. This technique will
give a novel advantage in the QCD perturbative analysis: enable to avoid
dealing  with the unmeasurable $x\rightarrow 0$ region.

\section{Truncated moments method within double logarithmic $ln^2x$ resummation
for the nonsinglet structure functions.}

Truncated moments of parton distributions have been lately used in the LO
and NLO DGLAP analysis \cite{b6}. Authors avoid in this way an
extrapolation of well known quark distributions behaviour to the
unmeasurable so unknown $x\rightarrow 0$ region. Apart from they receive
evolution equations for the truncated moments, in which $n$-th truncated
moment couples only with $n+j$-th ($j\geq 0$) moments and that the series of
couplings is convergent, what ensures a good accuracy. In our double
logarithmic analysis we use truncated moments of the unintegrated function
$f^{NS}(x,k^2)$ and for fixed coupling $\bar{\alpha_s}$ we can solve the
suitable equation analytically. Let us shortly explain this treatment. The
truncated $n$-th Mellin moment of the function $f^{NS}(x,k^2)$ from
(\ref{r2.13})-(\ref{r2.14}) is defined as:
\begin{equation}\label{r3.1}
\bar{f}^{NS}(x_0,n,k^2) \equiv \int\limits_{x_0}^1 dx x^{n-1} f^{NS}(x,k^2)
\end{equation}
The evolution equation (\ref{r2.9}), generating double logarithmic terms
$ln^2x$ in the truncated Mellin moment space takes a form:
\begin{eqnarray}\label{r3.2}
\bar{f}^{NS}(x_0,n,k^2) = \bar{f_0}^{NS}(x_0,n) + \bar{\alpha_s}
\int\limits_{k_0^2}^{k^2/x_0} \frac{dk'^2}{k'^2}
\nonumber \\
\times\int\limits_{x_0}^{1}dy y^{n-1} f^{NS}(y,k'^2)
\int\limits_{x_0/y}^{1} dz z^{n-1}\Theta \left(\frac{k^2}{k'^2}-z\right)
\nonumber \\
\end{eqnarray}
where $\Theta (t)$ is Heaviside's function:
\begin{equation}\label{r3.3}
\Theta(t)=\cases{1 & for~~ $t>0$ \cr 0 & for~~ $t\leq 0$ \cr}
\end{equation}
and we deal with the $x$-Bjorken region, where
\begin{equation}\label{r3.4}
x\geq x_0
\end{equation}
After taking into account the relations:
\begin{equation}\label{r3.5}
\int\limits_{x_0/y}^{1} dz z^{n-1}\Theta \left(\frac{k^2}{k'^2}-z\right) = 
\frac{1}{n} \left[\Theta (k^2-k'^2) + \Theta
(k'^2-k^2)\left(\frac{k^2}{k'^2}\right)^n -\frac{x_0^n}{y^n}\right];
~~~~~n\neq 0
\end{equation}
\begin{equation}\label{r3.6}
\int\limits_{x_0/y}^{1} dz z^{n-1}\Theta \left(\frac{k^2}{k'^2}-z\right) = 
\ln{\frac{y}{x_0}} + \Theta (k'^2-k^2)\ln{\frac{k^2}{k'^2}};~~~~~n=0
\end{equation}
one can obtain from (\ref{r3.2}):
\begin{eqnarray}\label{r3.7}
\bar{f}^{NS}(x_0,0,k^2) = \bar{f_0}^{NS}(x_0,0) + \bar{\alpha_s}
[\int\limits_{k^2}^{k^2/x_0} \frac{dk'^2}{k'^2} \ln{\frac{k^2}{k'^2}}
\bar{f}^{NS}(x_0,0,k'^2) 
\nonumber \\
+ \int\limits_{k_0^2}^{k^2/x_0}\frac{dk'^2}{k'^2}
\int\limits_{x_0}^{1} \frac{dy}{y} \ln{y} f^{NS}(y,k'^2)-\ln{x_0}\int
\limits_{k_0^2}^{k^2/x_0}\frac{dk'^2}{k'^2}\bar{f}^{NS}(x_0,0,k'^2)]
\nonumber \\
\end{eqnarray}
and
\begin{eqnarray}\label{r3.8}
\bar{f}^{NS}(x_0,n,k^2) = \bar{f_0}^{NS}(x_0,n) + \frac{\bar{\alpha_s}}{n}
[\int\limits_{k_0^2}^{k^2} \frac{dk'^2}{k'^2}\bar{f}^{NS}(x_0,n,k'^2) 
\nonumber \\
+ \int\limits_{k^2}^{k^2/x_0} \frac{dk'^2}{k'^2}\left(\frac{k^2}{k'^2}\right)^n
\bar{f}^{NS}(x_0,n,k'^2)- x_0^n\int\limits_{k_0^2}^{k^2/x_0}\frac{dk'^2}{k'^2}
\bar{f}^{NS}(x_0,0,k'^2)];~~~~~n\neq 0
\nonumber \\
\end{eqnarray}
For $n\neq 0$ we get the following solution (for details see Appendix B):
\begin{equation}\label{r3.9}
\bar{f}^{NS}(x_0,n,k^2) = 
\bar{f_0}^{NS}(x_0,n)\left(\frac{k^2}{k_0^2}\right)^{\gamma}\frac{R}{1+(R-1)x_0^n}
\end{equation}
where
\begin{equation}\label{r3.10}
R\equiv R(n,\bar{\alpha_s}) = \frac{n\gamma}{\bar{\alpha_s}}
\end{equation}
$\gamma$ is given in (\ref{r2.19}) and $\bar{f_0}^{NS}(x_0,n)$ is the
inhomogeneous term, independent on $k^2$:
\begin{equation}\label{r3.11}
\bar{f_0}^{NS}(x_0,n) = \int\limits_{x_0}^1 dx x^{n-1} f_0^{NS}(x)=
\frac{\bar{\alpha_s}}{n}\int\limits_{x_0}^1 \frac{dx}{x} (x^n-x_0^n) p_0(x)
\end{equation}
The input parton distribution $p_0(x)$ in the above formula denotes
$q_0^{NS}(x)$ for the unpolarized case or $\Delta q_0^{NS}(x)$ for the
polarized one, respectively. From (\ref{r3.9}) one can read that our
solution for the truncated moment $\bar{f}^{NS}(x_0,n,k^2)$ reduces to
(\ref{r2.18}) when $x_0=0$ what must be of course fulfilled.

\section{Some results for truncated moments $\bar{f}^{NS}(x_0,n,k^2)$, 
$\bar{F_2}^{NS}(x_0,n,k^2)$, $\bar{g_1}^{NS}(x_0,n,k^2)$ in the double 
logarithmic $ln^2x$ approximation.}

The truncated $n$-th moment of the unintegrated nonsinglet function
$f^{NS}(x,k^2)$ is given by eqs. (\ref{r3.9})-(\ref{r3.11}). In Figs.2-3 we
plot the moments of $f^{NS}(x,k^2)$ for different $n$ as a function of $x_0$
at $k^2=10 {\rm GeV}^2$. One can see that the ratio $p_f(x_0,n)$ defined as
\begin{equation}\label{r4.1}
p_f(x_0,n)\equiv\frac{\bar{f}^{NS}(x_0,n,k^2)}{\bar{f}^{NS}(0,n,k^2)}
\end{equation}
becomes very large ($\simeq 1$) at $x_0\approx 10^{-4}$
($p_f(x_0=10^{-4},n=1)=0.997$). This could be an advice that the truncated 
moments method is useful in the region $x\geq x_0\sim 10^{-4}$ because at lower 
$x$ the results for the full and the truncated moments are practically the same
(at least for the Regge input). From the other side double logarithmic $ln^2x$
terms from our approach become important in the small $x$ region $x\leq
10^{-2}$ \cite{b7}, \cite{b9}. So taking into account both above facts one should
choose the limit point $x_0$ in the truncated moments of $f^{NS}$, $g_1^{NS}$
or $F_2^{NS}$ as
\begin{equation}\label{r4.2}
10^{-2}\geq x_0 \geq 10^{-4}
\end{equation}
\begin{figure}[ht]
\begin{center}
\includegraphics[width=70mm]{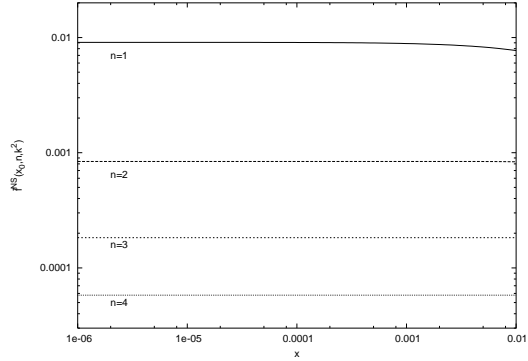}
\caption{Truncated Mellin moments $\bar{f}^{NS}(x_0,n,k^2)=
\int\limits_{x_0}^1 dx x^{n-1} f^{NS}(x,k^2)$ as a function of $x_0$ for
different $n$ at $k^2=10 {\rm GeV}^2$ in $ln^2x$ approach.}
\end{center}
\end{figure}
\begin{figure}[ht]
\begin{center}
\includegraphics[width=70mm]{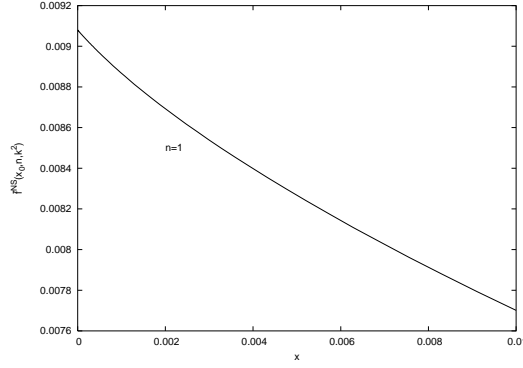}
\caption{The truncated Mellin moment $\bar{f}^{NS}
(x_0,n=1,k^2=10 {\rm GeV}^2)$, similarly as in Fig.2 but in a linear scale.}
\end{center}
\end{figure}
Truncated moments of the nonsinglet quark distributions are related to the
moments of $f^{NS}$ via:
\begin{equation}\label{r4.3}
\int\limits_{x_0}^1 dx x^{n-1} g_1^{NS}(x,k ^2)=
\int\limits_{x_0}^1 dx x^{n-1} g_1^{0NS}(x)+
\int\limits_{k_0^2}^{k^2/x_0} \frac{dk'^2}{k'^2(1+\frac{k'^2}{k^2})}
\bar{f}^{NS}(x_0,n,k'^2)
\end{equation}
for the spin dependent structure function and via:
\begin{equation}\label{r4.4}
\int\limits_{x_0}^1 dx x^{n-2} F_2^{NS}(x,k ^2)=
\int\limits_{x_0}^1 dx x^{n-2} F_2^{0NS}(x)+
\int\limits_{k_0^2}^{k^2/x_0} \frac{dk'^2}{k'^2(1+\frac{k'^2}{k^2})}
\bar{f}^{NS}(x_0,n,k'^2)
\end{equation}
for the unpolarized case. This together with (\ref{r3.9})-(\ref{r3.11})
gives the following formulae:
\begin{eqnarray}\label{r4.5}
I_1(x_0,n,k^2)\equiv\int\limits_{x_0}^1 dx x^{n-1} g_1^{NS}(x,k ^2)=
\bar{g_1}^{0NS}(x_0,n) 
\nonumber \\
+ B(x_0,n,k^2)[\bar{g_1}^{0NS}(x_0,n)-x_0^n\bar{g_1}^{0NS}(x_0,0)]
\nonumber\\
\end{eqnarray}
\begin{eqnarray}\label{r4.6}
I_2(x_0,n,k^2)\equiv\int\limits_{x_0}^1 dx x^{n-1} F_2^{NS}(x,k ^2)=
\bar{F_2}^{0NS}(x_0,n) 
\nonumber\\
+ B(x_0,n+1,k^2)[\bar{F_2}^{0NS}(x_0,n)-x_0^{n+1}\bar{F_2}^{0NS}(x_0,0)]
\nonumber\\
\end{eqnarray}
where
\begin{equation}\label{r4.7}
B(x_0,n,k^2) = \frac{\gamma (\frac{k^2}{k_0^2})^{\gamma}}{1+(R-1)x_0^n}
\int\limits_{\ln \frac{k_0^2}{k^2}}^{\ln \frac{1}{x_0}} dt 
\frac {e^{\gamma t}}{1+e^t}
\end{equation}
In Table I we collect results for different $n$-th moments of
$F_2^{NS}(x,k^2)$ and $g_1^{NS}(x,k^2)$ functions obtained from
(\ref{r4.5})-(\ref{r4.6}). The moments are truncated at $x_0 = 10^{-1},
10^{-2}, 10^{-3}$ and $10^{-4}$.
\begin{table}[ht]
\begin{center}
\begin{tabular}{|c|c|c|c|}
\hline\hline
$x_0$ & $n$ & $I_1(x_0,n,k^2)$ & $I_2(x_0,n,k^2)$ \\ \hline\hline
           & 1 & 0.144631 & 0.031977 \\
           & 2 & 0.039952 & 0.010018 \\
$10 ^{-1}$ & 3 & 0.014137 & 0.003945 \\
           & 4 & 0.006101 & 0.001843 \\ \hline
           & 1 & 0.219244 & 0.038540 \\
           & 2 & 0.043831 & 0.010413 \\
$10 ^{-2}$ & 3 & 0.014400 & 0.003975 \\
           & 4 & 0.006124 & 0.001846 \\ \hline
           & 1 & 0.229486 & 0.038791 \\
           & 2 & 0.043892 & 0.010416 \\
$10 ^{-3}$ & 3 & 0.014402 & 0.003975 \\
           & 4 & 0.006125 & 0.001847 \\ \hline
           & 1 & 0.230692 & 0.038801 \\
           & 2 & 0.043894 & 0.010416 \\
$10 ^{-4}$ & 3 & 0.014402 & 0.003975 \\
           & 4 & 0.006125 & 0.001847 \\
\hline\hline
\end{tabular}
\caption{\label{tab:par}Truncated moments $I_1(x_0,n,k^2)\equiv\int
\limits_{x_0}^1 dx x^{n-1} g_1^{NS}(x,k ^2)$ and $I_2(x_0,n,k^2)\equiv\int
\limits_{x_0}^1 dx x^{n-1} F_2^{NS}(x,k ^2)$ for different $x_0$ and $n$ in the 
$ln^2x$ approach.}
\end{center}
\end{table}
For all moments in Table I $k^2=10 {\rm GeV}^2$, $\alpha_s = 0.18$ and
\begin{equation}\label{r4.8}
\bar{F_2}^{0NS}(x_0,n)\equiv \int\limits_{x_0}^1 dx x^{n-1} F_2^{0NS}(x)
\end{equation}
\begin{equation}\label{r4.9}
\bar{g_1}^{0NS}(x_0,n)\equiv \int\limits_{x_0}^1 dx x^{n-1} g_1^{0NS}(x)
\end{equation}
Input parametrizations $g_1^{0NS}(x)$ and $F_2^{0NS}(x)$ are chosen at
$k_0^2=1 {\rm GeV}^2$ in the simple Regge form:
\begin{equation}\label{r4.10}
F_2^{0NS}(x) = \frac {35}{96} \sqrt{x} (1-x)^3
\end{equation}
\begin{equation}\label{r4.11}
g_1^{0NS}(x) = 0.838 (1-x)^3
\end{equation}
which satisfy the sea flavour symmetric Gottfried \cite{b10} and Bjorken 
\cite{b11} sum rules:
\begin{equation}\label{r4.12}
I_{GSR} = \int\limits_{0}^{1} \frac {dx}{x} F_2^{NS}(x,k ^2)= \frac {1}{3}
\end{equation}
\begin{equation}\label{r4.13}
I_{BSR} = \int\limits_{0}^{1} dx g_1^{NS}(x,k ^2)= \frac {1}{6}g_A = 0.21
\end{equation}
where $g_A=1.257$ is the axial vector coupling.
In order to determine the moment integrals we need knowledge of structure
functions over the entire region of $x$. The small $x$ behaviour of
structure functions is driven by the double logarithmic $ln^2x$ terms. This
$ln^2x$ approximation is however inaccurate in describing the $Q^2$
evolution for large values of $x$. Therefore double logarithmic $ln^2x$ 
approach should be completed by LO DGLAP $Q^2$ evolution. Dealing with 
truncated moments of $f^{NS}$ within unified $ln^2x+$LO resummation one 
encounters difficulties because suitable evolution equations are of course more 
complicated than in the pure double logarithmic approach here presented. 
Probably the only successful method to solve the evolution equations for 
truncated moments in $ln^2x+$LO treatment is the numerical calculus. We will 
discuss this in detail in the forthcoming paper.
Now we are able to employ the truncated moment method in calculation of the
contribution to moment integrals coming from the region of very small $x$.
Using the definition of the truncated moment (\ref{r3.1}) we can in a very
easy way find the double truncated moments of $f^{NS}$, $g_1^{NS}$ or 
$F_2^{NS}$:
\begin{equation}\label{r4.14}
\int\limits_{x_1}^{x_2} dx x^{n-1} f^{NS}(x,k^2) = 
\int\limits_{x_1}^{1} dx x^{n-1} f^{NS}(x,k^2) - 
\int\limits_{x_2}^{1} dx x^{n-1} f^{NS}(x,k^2)
\end{equation}
etc. In this way for the partial Bjorken sum rule, according to (\ref{r4.5}), 
we have: 
\begin{equation}\label{r4.15}
\int\limits_{x_1}^{x_2} dx g_1^{NS}(x,k^2=10) = I_1(x_1,1,10) - I_1(x_2,1,10)
\end{equation}
where $x_1$ and $x_2$ are both very small:
\begin{equation}\label{r4.16}
x_1 \leq x_2 \leq 10^{-2}
\end{equation}
We have thus estimated a contribution:
\begin{equation}\label{r4.17}
\Delta I_{BSR}(x_1,x_2,k^2) = \int\limits_{x_1}^{x_2} dx g_1^{NS}(x,k^2)
\end{equation}
and found that $\Delta I_{BSR}(10^{-4},10^{-2},10) = 0.011$, 
$\Delta I_{BSR}(10^{-5},10^{-3},10) = 0.0013$ what is equal to around $5\%$
and $0.6\%$ respectively of the value of the full sum (\ref{r4.13}). In our
above estimation we have assumed the simple Regge input parametrization of
$g_1^{0NS}$ (\ref{r4.11}). 

\section{Summary and conclusions.}

In our paper we have derived the integral equation for truncated moments of
the unintegrated nonsinglet structure function $f^{NS}(x,Q^2)$ in the case of
double logarithmic $ln^2x$ approximation. Analytical results at fixed
$\alpha_s$ for truncated moments of $F_2^{NS}$ and $g_1^{NS}$ have been
presented. We have received the clear, new solutions and what is important,
in the limit $x_0\rightarrow 0$ our calculations confirm the well
known analytical result for full moments of $f^{NS}(x,Q^2)$. The resummation of
$ln^2x$ terms at small $x$ goes beyond the standard LO (and even NLO) QCD
evolution of spin dependent parton densities. In the case of unpolarized
structure functions, the Regge behaviour, originating from the nonperturbative
contribution is the leading one at small $x$. Therefore the double
logarithmic approximation is very important particularly for the polarized
structure function $g_1$, which at low $x$ is dominated just by $ln^2x$
terms. In our paper we have obtained analytical solutions for truncated
Mellin moments of $g_1^{NS}$ (and $F_2^{NS}$ too). Dealing with truncated
moments at $x_0$: $\int\limits_{x_0}^{1} dx x^{n-1} g_1(x,Q^2)$ one can
avoid uncertainty from the unmeasurable very small $x\rightarrow 0$ region.
In this way the theoretical predictions of QCD analysis for structure
functions at small $x$ can be compared with experimental data without
necessity to extrapolate results into nonavailable $x<x_0$ range. In our
analysis for nonsinglet structure functions $g_1^{NS}$ and $F_2^{NS}$ we
have found their truncated at $x_0$ moments, what could be helpful in the
estimation of Bjorken ($g_1^{NS}$) and Gottfried ($F_2^{NS}$) sum rules. We
have estimated the contribution to the Bjorken sum rule from the very 
small $x$ region ($10^{-4}<x<10^{-2}$) and found it to be around $5\%$ of the
value of the sum. Evolution equations, we have derived for truncated Mellin 
moments of $f^{NS}(x,Q^2)$ generate correctly the leading small $x$ behaviour 
but do not describe $Q^2$ evolution at large values of $x$. In the integrals 
for moments of structure functions one need to have values of $g_1$ or $F_2$ 
from the entire range of $x$: $x_0\leq x \leq 1$ so for large $x$ one should 
include in the formalism LO Altarelli-Parisi (DGLAP) evolution. The aim of our 
paper was to focus attention on the truncated moments technique within $ln^2x$ 
approach itself. We found an analytical solution for the truncated moments
of structure functions within $ln^2x$ approximation. This is an important
stage before further investigations. The next step in the improvement of our 
analysis will be including LO DGLAP evolution into equations generating double 
logarithmic terms $ln^2x$ for truncated moments of the nonsinglet structure 
functions. This should give a proper determination of truncated and thus 
experimentally verifiable sum rules.

\section*{Acknowledgements}

The work presented here is the result of discussions of one of the
authors (D.K.) with Jan
Kwieci\'nski. He was, as usual, very helpful, patient
and stimulating. We are greatly indebted to him for many
years of teaching us QCD. It is painful to us to express
our appreciation of the help we received from  Jan Kwieci\'nski after he is
gone.

\appendix
\section{Analytical solution of the evolution equation for the full moments
of the function $f^{NS}(x,k^2)$ generating double logarithmic $\ln^2x$ effects 
at small $x$.}
  
After double differentiation of (\ref{r2.15}) with respect to $lnk^2$ one
obtains:
\begin{equation}\label{rA.1}
\frac{\partial^2 \bar{f}^{NS}(n,k^2)}{\partial t^2} = 
n \frac{\partial \bar{f}^{NS}(n,k^2)}{\partial t} - 
\bar{\alpha_s} \bar{f}^{NS}(n,k^2)
\end{equation}
with
\begin{equation}\label{rA.2}
t = \ln k^2
\end{equation}
Hence function $\bar{f}^{NS}(n,k^2)$ has a form:
\begin{equation}\label{rA.3}
\bar{f}^{NS}(n,k^2) = \left(\frac{k^2}{k_0^2}\right)^{\gamma} 
H(n,\bar{\alpha_s})
\end{equation}
where $H(n,\bar{\alpha_s})$ from initial conditions should be:
\begin{equation}\label{rA.4}
H(n,\bar{\alpha_s}) = \bar{f_0}^{NS}(n) R(n,\bar{\alpha_s})
\end{equation}
and $\gamma$ is given by (\ref{r2.19})-(\ref{r2.20}). After inserting the
solution (\ref{rA.3})-(\ref{rA.4}) into (\ref{r2.15}) one finds that $R$ is
connected with $\bar{\alpha_s}$, $n$ and $\gamma$ via:
\begin{equation}\label{rA.5}
R(n,\bar{\alpha_s}) = \frac{n\gamma}{\bar{\alpha_s}}
\end{equation}
Finally, the solutions (\ref{rA.3})-(\ref{rA.5}) for $n$-th moment 
$\bar{f}^{NS}(n,k^2)$ giving (\ref{r2.18}), imply via inverse Mellin
transform
\begin{equation}\label{rA.6}
f^{NS}(x,k^2) = \frac{1}{2\pi i}\int\limits_{c-i\infty}^{c+i\infty}
dn x^{-n} \bar{f}^{NS}(n,k^2)
\end{equation}
the following small $x$ behaviour for $f^{NS}(x,k^2)$:
\begin{equation}\label{rA.7}
f^{NS}(x,\tau) \sim \frac{\sqrt{2n_0}}{4\bar{\alpha_s}\sqrt{\pi}}
(n_0 \ln\tau + 1) (\frac{\tau}{x})^{n_0}(\ln\frac{\tau}{x})^{-3/2}
\end{equation}
where
\begin{equation}\label{rA.8}
\tau = \left(\frac{k^2}{k_0^2}\right)^{1/2}
\end{equation}
and $n_0$ defined in (\ref{r2.20}) denotes branch point singularity of the
anomalous dimension $\gamma$.

\section{Analytical solution of the evolution equation for truncated 
moments of the function $f^{NS}(x,k^2)$ generating double logarithmic $\ln^2x$ 
effects at small $x$.}

From the definition of the $n$-th ($n\neq 0$) truncated moment of $f(x,k^2)$ 
(\ref{r3.1}) one can read the following relation:
\begin{equation}\label{rB.1}
\frac{\partial \bar{f}^{NS}(x_0,n,k^2)}{\partial x_0} = 
x_0^n \frac{\partial \bar{f}^{NS}(x_0,0,k^2)}{\partial x_0}
\end{equation}
On the strength of this above relation we can replace $\bar{f}^{NS}(x_0,0,k'^2)$
in (\ref{r3.8}) via:
\begin{equation}\label{rB.2}
\bar{f}^{NS}(x_0,0,k'^2) = x_0^{-n} \bar{f}^{NS}(x_0,n,k'^2) - 
n\int\limits_{x_0}^{1}\frac{dy}{y^{n+1}}\bar{f}^{NS}(y,n,k'^2)
\end{equation}
and hence we get for (\ref{r3.8}):
\begin{eqnarray}\label{rB.3}
\bar{f}^{NS}(x_0,n,k^2) = \bar{f_0}^{NS}(x_0,n) + \frac{\bar{\alpha_s}}{n}
\int\limits_{k^2}^{k^2/x_0}
\frac{dk'^2}{k'^2}\left[\left(\frac{k^2}{k'^2}\right)^n - 1\right]
\bar{f}^{NS}(x_0,n,k'^2) 
\nonumber \\
+ \bar{\alpha_s}x_0^{n}\int\limits_{k^2}^{k^2/x_0} \frac{dk'^2}{k'^2}
\int\limits_{x_0}^{1}\frac{dy}{y^{n+1}}\bar{f}^{NS}(y,n,k'^2)
\nonumber \\
\end{eqnarray}
Now we "guess" the solution of (\ref{r3.8}) in the form:
\begin{equation}\label{rB.4}
\bar{f}^{NS}(x_0,n,k^2) = 
\bar{f_0}^{NS}(x_0,n)\left(\frac{k^2}{k_0^2}\right)^{\gamma} 
g(x_0,n,\bar{\alpha_s})
\end{equation}
where we postulate the same anomalous dimension $\gamma$ to be in agreement
with the result for the full moment:
\begin{equation}\label{rB.5}
\bar{f}^{NS}(x_0=0,n,k^2) = \bar{f}^{NS}(n,k^2)
\end{equation}
After insertion (\ref{rB.4}) into (\ref{rB.3}) we can find a simple form of
the auxiliary function $g(x_0,n,\bar{\alpha_s})$:
\begin{equation}\label{rB.6}
g(x_0,n,\bar{\alpha_s}) = \frac{n}{n-\gamma + \gamma x_0^n}
\end{equation}
or equivalently:
\begin{equation}\label{rB.7}
g(x_0,n,\bar{\alpha_s}) = \frac{R}{1+(R-1)x_0^n}
\end{equation}
with $R=R(n,\bar{\alpha_s})$ given by (\ref{r3.10}).

\end{document}